\documentclass[9pt,twocolumn,twoside]{osajnl}
\usepackage{upgreek}
\usepackage{units}
\usepackage{graphicx}

\journal{ol} % Choose journal (ao, aop, josaa, josab, ol, pr)

\setboolean{shortarticle}{true}
\title{Generalized axicon-based generation of nondiffracting beams}

\author[1]{Keyou Chen}
\author[1,2]{Michael Jenne}
\author[3]{Daniel Günther Grossmann}
\author[1,*]{Daniel Flamm}
\affil[1]{TRUMPF Laser- und Systemtechnik GmbH, Johann-Maus-Strasse 2, 71254 Ditzingen, Germany}
\affil[2]{Institute of Applied Physics, Abbe Center of Photonics, Friedrich Schiller University Jena, Albert-Einstein-Strasse 15, 07745 Jena, Germany}
\affil[3]{TRUMPF Laser GmbH, Aichhalder Strasse 39, 78713 Schramberg }

\affil[*]{Corresponding author: daniel.flamm@trumpf.com}

\dates{Submitted \today}

\ociscodes{(140.3300) Laser beam shaping; (050.1970) Diffractive optics; (090.1995) Digital holography; (140.3390) Laser materials processing.}

\doi{}

\begin{abstract}
We generalize the well-known method of generating nondiffracting beams based on axicons by allowing phase modulations with azimuthal dependencies. This generalization includes the description of Bessel-like beams of zero order, higher orders and superpositions thereof. We present the enormous benefit of our approach for highly efficient and robust shaping of nondiffracting beams with arbitrary transverse profiles. The concept's versatility is demonstrated by discussing generation and propagation of various nondiffracting beams with potential regarding laser materials processing.
\end{abstract}

\setboolean{displaycopyright}{false}

\begin{document}

\maketitle

\noindent
Beside most diverse fields of application in research and science, see e.g. \cite{mcgloin2003, fahrbach2010, zhang2018}, we encounter nondiffracting beams even in our every-day life. Bar code scanners with built-in axicons make it easy for cashiers to register our goods \cite{marom1994}. The resulting spot with dimensions similar to the bar code features remains substantially constant along a remarkable distance between scanning head and the code identifying the product \cite{marom1994}. It seems that this radiation no longer obeys the rules of diffraction and, thus, its inventors Durnin \textit{et al.}\ called this class of wave fields ``nondiffracting'' or ``diffraction-free'' \cite{durnin1987}. Equally used is the term ``Bessel-like'', since the transverse field distribution equals Bessel functions \cite{durnin1987b}. A few years later, first efforts were made to alter the transverse field properties of the zero-order Bessel beam while maintaining its nondiffracting properties. It was the birth of pure higher-order Bessel-like beams \cite{lee1994} and beams being superpositions thereof \cite{vasara1989}. Later, more detailed studies were carried out to find further propagation-invariant solutions to the Helmholtz equation defined in different coordinate systems \cite{woerdemann2012}. The class of Mathieu beams \cite{gutierrez2000, woerdemann2012}, for example, already shows a multitude of possible transverse field distributions. The present work takes up exactly this point and provides a simple and comprehensive recipe to generate nondiffracting beams with an enormous diversity regarding the transverse intensity profile which includes above mentioned cases and goes far beyond, respectively. Since these beams are intended to serve as subtle tools for materials processing, we place particular emphasize to maximum efficiency and power resistance of a single beam shaping element. 
\par
\begin{figure}[t]
\centering
\includegraphics[draft=false, width=1.0\linewidth]{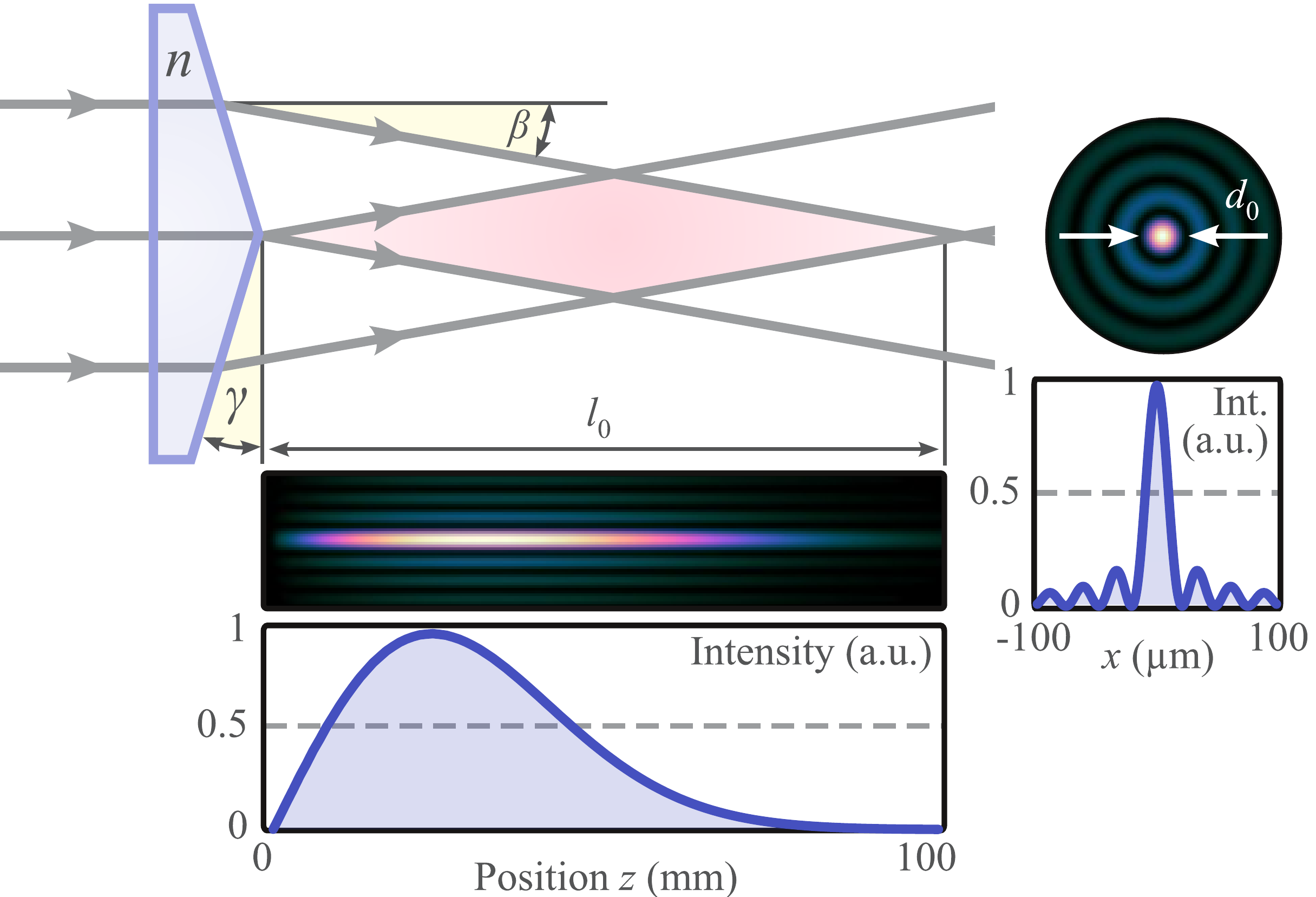}
\caption{Schematic representation of shaping a Bessel-Gaussian beam of zero order using a refractive axicon defined by refractive index $n$ and axicon opening angle $\gamma$ \cite{mcgloin2005}. }
\label{fig:Axi}\vspace{-0.1cm}
\end{figure}

 
Although we see a plethora of applications in the field of microscopy and particle manipulation, our main motive is the use of this class of beams for ultrafast processing of transparent materials \cite{flamm2015, flamm2019, jenne2018}. Very recently, there is ongoing research for advanced cutting of glasses using asymmetric Bessel-like beams to control the formation of cracks. Established concepts make use of spatial frequency filtering or controlled aberrations \cite{hendricks2016, meyer2017, dudutis2018} but come at the expense of significant power loss or reduced process stability. As will be shown in the following, the concept presented also includes such special cases at which radial symmetry of the transverse optical field is intentionally broken. In fact, the shaped transverse intensity distribution may exhibit neither radial nor point symmetry. At the same time, all outstanding properties of nondiffracting beams remain, such as self-healing \cite{mcgloin2005}, natural resistance to spherical aberrations \cite{flamm2019}, extreme aspect ratio of longitudinal to transverse dimensions ($l_0 / d_0 > 1000$, see Fig.\,\ref{fig:Axi}) \cite{mcgloin2005}, or easy and efficient generation \cite{mcgloin2005}. 
The later property is achieved in particular by using an axicon as central beam shaping element. An axicon is a conically ground lens \cite{mcleod1954} and, by design, completely defined by axicon opening angle $\gamma$ and refractive index $n$, cf.\,Fig.\,\ref{fig:Axi} \cite{mcgloin2005}. If illuminated by a plane wave with a sufficient degree of coherence, a special interference pattern will be observed starting directly behind the axicon tip that is characterized by an elongated intensity maximum on the optical axis surrounded by weaker, equally spaced rings. The peak intensity ratio of the first ring to the central maximum is about $0.17$, see Fig.\,\ref{fig:Axi}. This property is particularly favorable for nonlinear absorption processes since there is a large intensity- and, thus, pulse energy range where the processing beam modifies the material only at the central maximum but not at the side maxima. The mentioned interference pattern is a consequence of field components propagating along a cone towards the optical axis with refraction angle $\beta \approx \left(n-1\right)\gamma$ and corresponding radial component of the wavevector $k_r = k_0\beta$ (in thin element approximation) \cite{mcgloin2005, leach2006}. \par
The zero-order Bessel-like beam thus generated belongs to the class of nondiffracting beams since the optical field $u\left(x,y,z\right)$ fulfills Durnin's fundamental plane wave ansatz \cite{durnin1987b}
\begin{align}\label{eq:1} \nonumber
u\left(x,y,z\right)= &\exp{\left(\imath k_z z\right)}\int_0^{2\uppi}{\mathrm{d}\phi\,a\left(\phi\right)}   \\ 
& \times \exp{\left[\imath k_t\left(x\cos{\phi} + y\sin{\phi}\right)\right]}.
\end{align}
Here, $a\left(\phi\right) = A\left(\phi\right)\exp{\left[\imath \Theta\left(\phi\right)\right]}$ is an arbitrary complex-valued function depending on the angular coordinate $\phi$ and $k_t$, $k_z$ are the transverse and longitudinal wave vector components, respectively \cite{durnin1987, gutierrez2000}. For the described axicon-based generation $k_t$ equals the radial component of the wave vector $k_r$ and $a\left(\phi\right) = 1$. For this particular case, the integrand of Eq.\,(\ref{eq:1}) simplifies into $\exp{\left[\imath \Phi^{\text{axi}}\left(r\right)\right]}$ with the known phase modulation of an ideal thin axicon $\Phi^{\text{axi}}\left(r\right) = k_rr$.
\par 
Techniques to generate further optical fields that fulfill Eq.\,(\ref{eq:1}) are available in a variety of designs, see e.g., \cite{cottrell2007nondiffracting, lopez2010shaped, lopez2010method}. The first nondiffracting beams were generated by ring-slit apertures (far-field generation) \cite{durnin1987}. Clearly more efficient are (near-field) concepts based on axicons \cite{mcgloin2005}. Combinations thereof were equally discussed as their flexible holographic realisation using liquid-crystal displays \cite{woerdemann2012}. Since the concept may serve to develop novel material processing strategies using ultrashort pulsed lasers with comparatively ``expensive'' photons, we are looking for a particularly efficient beam shaping technique and only allow phase modulations caused by a single axicon-like element illuminated by a fundamental Gaussian beam. 
\begin{figure*}[ht]
\centering
\includegraphics[draft=false, width=1.0\linewidth]{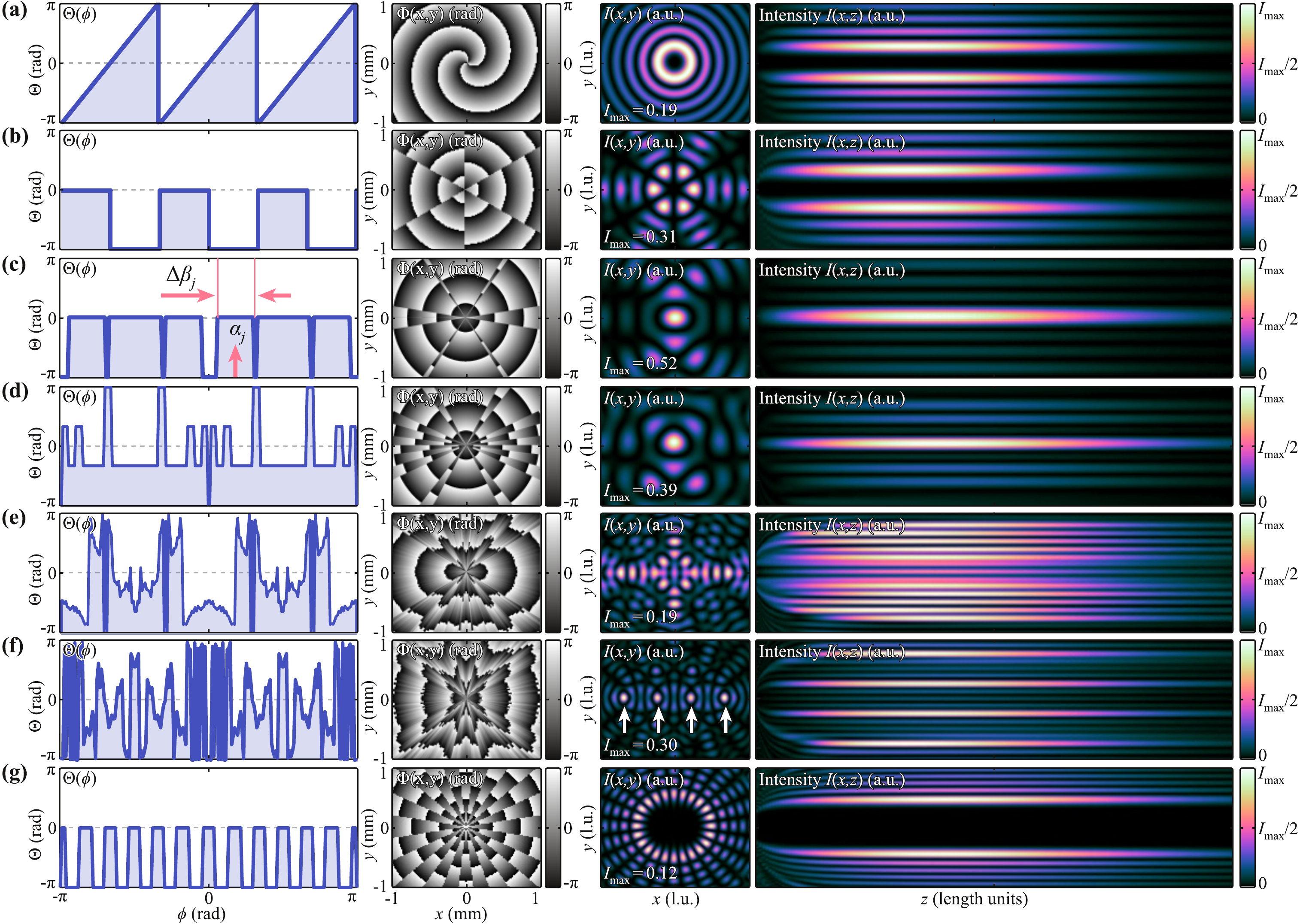}
\caption{Generation and propagation behaviour of a selection of generalized nondiffracting beams. Each row depicts the azimuthal phase dependency $\Theta\left(\phi\right)$, the corresponding, axicon-like phase modulation $\Phi\left(x,y\right)$ (both in a modulo-$2\uppi$ representation) and the resulting shaped intensity profile as transverse intensity profile $I\left(x,y\right)\vert_{z=z'} \equiv I\left(x,y\right)$ and as longitudinal intensity cross section $I\left(x,z\right)\vert_{y=0} \equiv I\left(x,z\right)$, respectively (from left to right). (a) Higher-order Bessel-like beam $\ell = 3$. (b) Superposition of higher-order Bessel-like beams. (c) Elliptical Bessel-like beam. (d) Asymmetric Bessel-like beam, exhibiting neither radial -- nor point symmetry. (e) Cross-like Bessel-like beam. (f) Fourfold split Bessel-like beam. (g) Mathieu-like beam of order $12$ (even) \cite{woerdemann2012}. In all cases a fundamental Gaussian beam is illuminating the phase modulation $\Phi\left(x,y\right)$. Thus, all shown nondiffracting beams are Bessel-Gaussian-like. The colormap representation is created based on the cubehelix concept introduced by Green \cite{green2011colour}.}
\label{fig:bess}\vspace{-0.2cm}
\end{figure*}
Setting $A\left(\phi\right) = 1$ (no amplitude modulation) simplifies the integrand in Eq.\,(\ref{eq:1}) to $\exp{\left[\imath \Phi\left(r, \phi\right)\right]}$ with the generalized phase modulation of axicon-like elements
\begin{align}\label{eq:2}
    \Phi\left(r,\phi\right) = k_rr+\Theta\left(\phi\right).
\end{align}
This generalized approach maintains the radial slope of an axicon $\partial\Phi\left(r,\phi\right)/\partial r=k_r$ and, additionally, allows azimuthal dependencies for the phase modulation. In principle, no restrictions have to be applied to $\Theta\left(\phi\right)$. Continuous functions are as conceivable as those containing discrete jumps. This general approach includes several well-known special cases. Setting $\Theta\left(\phi\right)=\ell\phi$ with $\ell \in \mathbb{Z}$ results in holograms to shape Bessel-like beams of higher-order, see Fig.\,\ref{fig:bess}\,(a). In this particular case $\ell=3$ generates a nondiffracting beam with an on-axis phase singularity being accompanied with a distinct intensity minimum \cite{lee1994}. Also included are phase modulations generating superpositions of higher-order Bessel-like beams \cite{vasilyeu2009}, such as, e.g., $\Theta\left(\phi\right)=\arg{\left[\exp{\left(\imath \ell \phi\right)} + \exp{\left(-\imath \ell \phi\right)}\right]}$, $\ell \in \mathbb{Z}$ shaping so-called ``petal-like'' beams with nondiffracting behaviour \cite{vasilyeu2009}. For this particular case $\Theta\left(\phi\right)$ exhibits several discrete $\uppi$-phase jumps as depicted in Fig.\,\ref{fig:bess}\,(b). \par 
The challenge now is to define the exact dependency of $\Theta\left(\phi\right)$ that will meet the requirements for a desired transverse intensity profile of a nondiffracting beam. In contrast to the two previous cases [cf. Figs.\,\ref{fig:bess}\,(a) and (b)] where $\Theta\left(\phi\right)$ was set analytically, an iterative approach is chosen. However, we would like to emphasize that the procedure described for determining $\Phi\left(r,\phi\right)$ represents only one of many possible approaches. Our idea is to implement phase disturbances into $\Phi\left(r,\phi\right)$ to ensure that the ideal on-axis interference condition, shaping the original Bessel-like beam of zero order no longer exists. This modified interference situation around the optical axis allows to shape the transverse intensity profile in a controlled manner. For this purpose we implement a number of $j_{\text{max}}$ phase disturbances by defining angular segments of position $\alpha_j$, width $\Delta\beta_j$ and constant phase offset $\Theta_j$. As depicted by the selected example of Fig.\,\ref{fig:bess}\,(c), $\Theta\left(\phi\right)$ is completely determined by the set of parameters $\left\{ \alpha_j, \Delta\beta_j, \Theta_j \right\}$. The optimization algorithm propagates the input optical field $u\left(r,\phi\right) = U\left(r,\phi\right)\exp{\left[\imath\Phi\left(r,\phi\right)\right]}$ to a well defined propagation distance $z=z'=\text{const.}$ where the Bessel-like beam exists. Due to the propagation invariance, exact knowledge about $z'$ does not play a decisive role. The amplitude distribution $U\left(r,\phi\right)$ might be a scalar plane wave, a super-Gaussian function or as in the following examples a fundamental Gaussian beam with a plane phase. An adapted merit function determines the deviations of the actual transverse intensity profile $I\left(x,z\right)\vert_{y=0}$ and a target intensity distribution. Depending on the respective application this might be, e.g., an elliptical or a cross-like transverse beam profile $I\left(x,y\right)$, see Figs.\,\ref{fig:bess}\,(c) and (e). Equivalently conceivable is to define the merit function as the sum of intensity differences at a few well defined pixels, see white arrows denoting distinct maxima in $I\left(x,z\right)\vert_{y=0}$ of Fig.\,\ref{fig:bess}\,(f). The actual optimization process, thus, the iterative search for phase offsets $\Theta_j$ in the defined angular segments, which minimizes the merit function is done by, e.g., a stochastic optimization method \cite{lagarias1998}. \par
In practice, we found that dividing $\Theta\left(\phi\right)$ into a maximum number of $200$ fan segments of equal angular width $\Delta\beta_j = 2\uppi / 200$ results in both an efficient design strategy and an enormous diversity of focus shapes. Since most of the targeted nondiffractive beams exhibit symmetry to either $x$- and/or $y$-axis we add point-symmetry restriction to the azimuthal phase dependency $\Theta\left(\phi+\uppi\right) = \Theta\left(\phi\right)$ and, thus, reduce the maximum number of free parameters for the optimization routine to $50$. It should be noticed at this point, however, that the optimal choice of parameters for the optimization ($j_{\text{max}}, \Delta\beta_j$) depends on several factors such as, e.g., target intensity distribution or resolution and dynamical range of the beam shaping element. All nondiffracting beam profiles and corresponding phase masks $\Phi\left(r,\phi\right)$ discussed in the following are designed to be realized with a conventional liquid-crystal phase-only spatial light modulator (SLM) with pixels of $>\unit[10]{\upmu m}$ pitch. For examples, see $\Phi\left(r,\phi\right)$ in Fig.\,\ref{fig:bess} (in modulo-$2\uppi$ representation).
\par 
Figure \ref{fig:bess} shows selected examples of nondiffracting beams with tailored transverse intensity profiles illustrating the broad spectrum of beam shapes included in our generalized concept. In each row (a) -- (g) we depict the azimuthal phase dependency $\Theta\left(\phi\right)$, the corresponding axicon-like phase modulation $\Phi\left(x,y\right)$ defined by Eq.\,(\ref{eq:1}) (for convenience plotted in Cartesian coordinates) and the resulting transverse and longitudinal intensity profile $I\left(x,y\right)\vert_{z=z'} \equiv I\left(x,y\right)$ and $I\left(x,z\right)\vert_{y=0} \equiv I\left(x,z\right)$, respectively. The well-known and already discussed cases of generating Bessel-like beams of higher-order and superposition thereof are shown as examples in Figs.\,\ref{fig:bess}\,(a) and (b). \par 
The generation and propagation characteristic of a nondiffracting beam with elliptical on-axis intensity maximum is depicted in Fig.\,\ref{fig:bess}\,(c). Here, $\Theta\left(\phi\right)$ is characterized by several angular segments with phase differences of $\uppi$. These phase shifts distort the on-axis interference and generate an elliptical central maximum with an aspect ratio of long to short axis of $1.5$. The peak intensity ratio of the first side lobe to the central maximum is about $0.4$ and, thus, there is still enough room to control nonlinear absorption processes. To demonstrate high efficiency of our beam shaping process, we determine the peak intensity ratio of this elliptical Bessel beam  to the Bessel-like beam of zero order (cf. Fig.\,\ref{fig:Axi}) to $I_{\text{max}}=0.52$. This example of a nondiffracting beam represents a promising candidate for facilitated single-pass cleaving of glasses using ultrashort pulsed lasers. Asymmetric beam profiles can be used to create well-orientated cracks inside the processing volume resulting in a significant reduced force necessary to separate the sample. A detailed analysis of corresponding laser-matter interaction including processing results is already submitted for publication by Jenne \textit{et al.} \cite{jenne2019}. \par 
The next remarkable example is depicted in Fig.\,\ref{fig:bess}\,(d) where a beam with clearly nondiffracting propagation properties and corresponding transverse intensity profile $I\left(x,y\right)$ is presented that shows neither radial nor point symmetry. The required azimuthal phase function already exhibits several fan segments with three discrete phase values. \par 
The axicon-like phase mask defined by $\Theta\left(\phi\right)$ and depicted in Fig.\,\ref{fig:bess}\,(e) already contains high-frequency components but is still continuous (please note the modulo-$2\uppi$ representation). The Bessel-like beam generated in this way exhibits a cross-like structure as transverse intensity profile. Such an example could be useful for certain applications connected to selective etching of ultrafast laser modified transparent materials. Here, modifications in all spatial directions become possible with a single pulse that enable highly efficient processing and etching of large volumes. \par 
Concepts to split laser radiation into several copies are equally relevant for strategies connected to particle manipulation and materials processing. After all, throughput is often to be maximized or the full performance of the laser system regarding power or energy is to be exploited. One suitable candidate which is particularly interesting for reasons of efficiency can be seen in Fig.\,\ref{fig:bess}\,(f). Here, four clearly separated maxima in $I\left(x,y\right)$ each with relative values of $0.3$ with respect to the maximum of a zero-order Bessel-like beam being normalized to unit intensity (cf. Fig.\,\ref{fig:Axi}) demonstrate a highly efficient beam shaping concept.
\par
Finally, in Fig.\,\ref{fig:bess}\,(g) we present the generation and intensity distribution of a nondiffracting beam that is, in principle, identical to a pure even Mathieu beam of order $m=12$. One could assume that, as in the case discussed in Fig.\,\ref{fig:bess}\,(b), there are $\uppi$-phase jumps of equal periodicity that determine $\Theta\left(\phi\right)$ and, in fact, the phase differences are exactly $\uppi$, however, the periods, on the other hand, differ minimally. As mentioned in the introduction, the class of Mathieu beams represent propagation invariant solutions of the Helmholtz equation in elliptic cylindrical coordinates \cite{gutierrez2000, woerdemann2012}. Nevertheless, several of them can be generated by the introduced generalized axicons -- a further example proving the remarkable diversity of our beam shaping approach. In principle, our generalized axicons do not modulate amplitudes by setting $a\left(\phi\right) = \exp{\left[\imath \Theta\left(\phi\right)\right]}$, cf.\,Eq.\,(\ref{eq:1}), such Mathieu beams can be generated that do not require a significant amplitude filtering. Such cases can be identified by extended intensity minima occurring in the corresponding far-field ring \cite{woerdemann2012}. Although we insist on maximum efficiency in this study, allowing amplitude \textit{and} phase modulations in $a\left(\phi\right)$, cf.\,Eq.\,(\ref{eq:1}), could lead to further degrees of freedom in the generation of tailored nondiffractive beams. However, in case of using an SLM as central beam shaping element, this would require complex-amplitude-to-phase-only coding techniques \cite{arrizon2007}.
\par 
Please note that we intendedly omit clear length specifications to all plotted intensity distributions $I\left(x,y\right)$ and $I\left(x,z\right)$, respectively, of Fig.\,\ref{fig:bess} for the following reason. The transverse and longitudinal dimensions of the nondiffracting beam (cf. Fig.\,\ref{fig:Axi}) can easily be controlled by raw-beam diameter, axicon angle $\gamma$ or a subsequent telescopic magnification \cite{mcgloin2005,jenne2018, flamm2019} and can easily exceed several orders of magnitude. For example, the total length of the $z$-axes of the $I\left(x,z\right)$-plots of Figs.\,\ref{fig:bess}\,(a) -- (g) could equally be $\unit[100]{\upmu m}$ or $\unit[100]{mm}$. 
\par
The concept presented can partially be applied to so called second types of nondiffracting beams, too, \cite{woerdemann2012, baumgartl2008} such as the interesting case of Bessel-like beams propagating on accelerating trajectories \cite{chremmos2012}.
\par
The experimental generation of Bessel-like beams, as well as higher-order versions or superpositions thereof has already been presented by diffractive optical elements or digital holography \cite{vasilyeu2009, bergner2018, flamm2019}. We therefore see no difficulties to realize the generalized nondiffracting beams presented in this work in the same way, show simulations only, and refer to the work of Jenne \textit{et al.} \cite{jenne2019} for experimental verification based on SLMs. Equally conceiveable would be the realisation of discussed phase modulations using refractive axicon-like elements fabricated by free form techniques. Reference \cite{jenne2019} will additionally demonstrate the impact of the presented nondiffracting beams during processing transparent materials using time resolved microscopy \cite{bergner2018, jenne2018multi}.\par 
To conclude, we presented a remarkably simple, user-friendly and efficient concept to design and generate nondiffracting beams with tailored transverse intensity profiles. Our general approach expands the axicon-like generation of Bessel-like beams by introducing an arbitrary azimuthal dependency $\Theta\left(\phi\right)$, cf. Eq.\,(\ref{eq:2}). Based on this definition we demonstrated the tremendous variety of possible nondiffracting beam shapes by means of selected examples and discussed their beneficial use in the context of laser materials processing.

%\section*{Acknowledgement.}
%We thank Charlotte Peschke, TRUMPF GmbH\,+\,Co.\,KG, for creating our visualizations.

\section*{Disclosures.}
The authors declare no conflicts of interest.

% Bibliography
\bibliography{sample}

% Full bibliography added automatically for Optics Letters submissions; the following line will simply be ignored if submitting to other journals.
% Note that this extra page will not count against page length
%\bibliographyfullrefs{sample}

%Manual citation list
%\begin{thebibliography}{1}
%\bibitem{Zhang:14}
%Y.~Zhang, S.~Qiao, L.~Sun, Q.~W. Shi, W.~Huang, %L.~Li, and Z.~Yang,
 % \enquote{Photoinduced active terahertz metamaterials with nanostructured
  %vanadium dioxide film deposited by sol-gel method,} Opt. Express \textbf{22},
  %11070--11078 (2014).
%\end{thebibliography}

\end{document}